\documentclass[twocolumn,twoside,preprintnumbers]{revtex4}%
\usepackage{amsfonts}
\usepackage{epsfig}
\usepackage{graphicx}
\usepackage{amsmath}
\usepackage{amssymb}%
\begin{document}
\title{Dynamical phase transition in vibrational surface modes}
\author{H. L. Calvo}
\email{calvoh@famaf.unc.edu.ar}
\affiliation{Facultad de Matem\'{a}tica, Astronom\'{\i}a y F\'{\i}sica, Universidad
Nacional de C\'{o}rdoba, Ciudad Universitaria, 5000 C\'{o}rdoba, Argentina.}
\author{H. M. Pastawski}
\email{horacio@famaf.unc.edu.ar}
\affiliation{Facultad de Matem\'{a}tica, Astronom\'{\i}a y F\'{\i}sica, Universidad
Nacional de C\'{o}rdoba, Ciudad Universitaria, 5000 C\'{o}rdoba, Argentina.}

\begin{abstract}
We consider the dynamical properties of a simple model of vibrational surface
modes. We obtain the exact spectrum of surface excitations and discuss their
dynamical features. In addition to the usually discussed localized and
oscillatory regimes we also find a second phase transition where the surface
mode frequency becomes purely imaginary and describes an overdamped regime.
Noticeably, this transition has an exact correspondence to the oscillatory -
overdamped transition of the standard oscillator with a frictional force
proportional to velocity.

\end{abstract}
\maketitle

\baselineskip=12pt

\section{Introduction}

\begin{figure}
[h]
\begin{center}
\includegraphics*[
height=3.0445cm,
width=7.0973cm
]%
{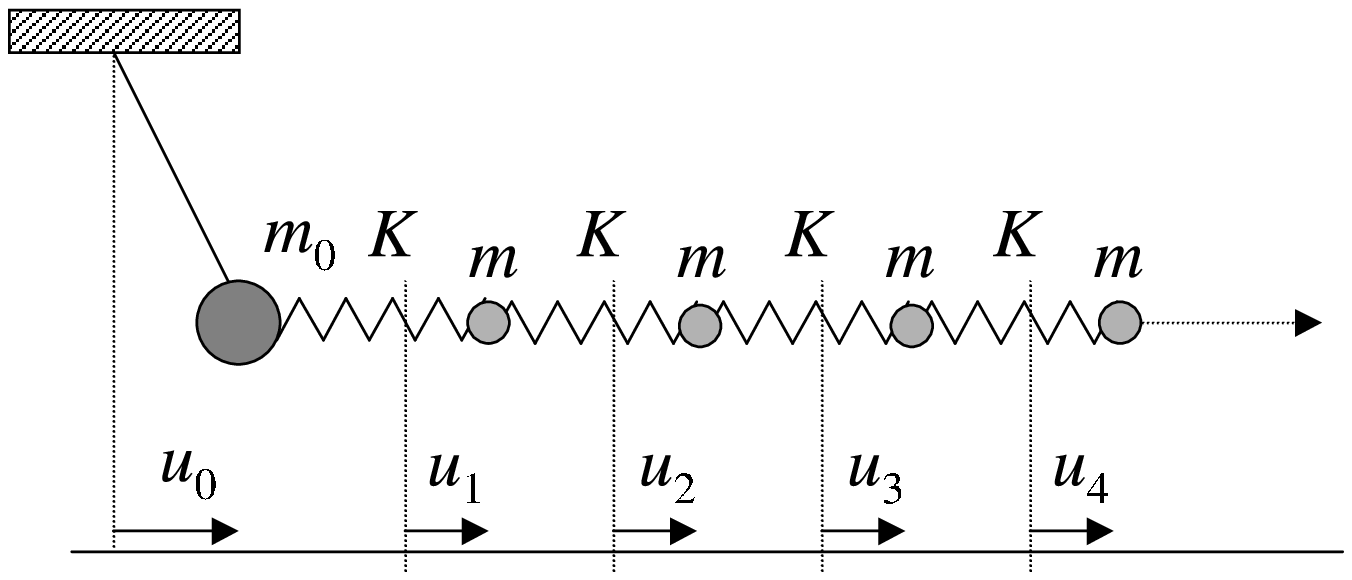}%
\caption{Scheme of the model: a simple pendulum (surface oscillator) is
coupled to the bulk masses.}%
\label{fig_scheme}%
\end{center}
\end{figure}

In a classical oscillator, dissipation is described considering an equation of
motion with a friction term proportional to the velocity \cite{Feynman}.
Ignoring microscopic details, all environmental effects are summarized
phenomenologically in the coefficient $\eta_{0}$ of this term. We propose here
a simple and time-reversal invariant model whose analytical solution yield
dissipation in the thermodynamic limit. We consider a variation of the Rubin
model \cite{Ingold, Rubin} shown in Fig.\ref{fig_scheme}: a "surface"
oscillator (represented as a pendulum) with natural frequency $\omega_{0}%
$\ and mass $m_{0}$ coupled to an ordered and semi-infinite chain of\ "bulk"
oscillators whose masses are $m$. As occurs with a \textquotedblleft Brownian
bath\textquotedblright\ \cite{Hanggi-Ingold}, Ohmic dissipation will
require\ $\alpha=m/m_{0}\ll1$.%

The equations of motion are,%
\begin{equation}%
\begin{array}
[c]{l}%
\left[  \dfrac{\mathrm{d}^{2}}{\mathrm{d}t^{2}}+\omega_{0}^{2}+\alpha
\omega_{x}^{2}\right]  u_{0}(t)-\alpha\omega_{x}^{2}u_{1}(t)=0,\\
\left[  \dfrac{\mathrm{d}^{2}}{\mathrm{d}t^{2}}-2\omega_{x}^{2}\right]
u_{n}(t)-\omega_{x}^{2}\left[  u_{n-1}(t)+u_{n+1}(t)\right]  =0,
\end{array}
\label{eq_motion}%
\end{equation}
where $\omega_{x}=\sqrt{K/m}$ is the exchange frequency between neighbor
oscillators and $u_{n}(t)$ denotes the $n$th oscillator displacement from
equilibrium. Assuming that the number of bulk oscillators is finite, the
time-reversal invariance present in these equations is obvious. We will show
below how irreversibility appears in the thermodynamic limit where the number
bulk oscillators become infinite.

\section{Frequency Domain}

The equations of motion described above can be solved by Fourier transforming
the displacement:%
\begin{equation}
u_{n}(t)=\frac{1}{2\pi}\int_{-\infty}^{\infty}\mathrm{d}\omega e^{-\mathrm{i}%
\omega t}u_{n}(\omega),
\end{equation}
and replacing it in Eq.(\ref{eq_motion}):%
\begin{equation}
\left(  \omega^{2}\mathbb{I-M}\right)  \mathbf{u}(\omega)=\mathbf{0}.
\end{equation}
This equation is solved by the eigenfrequencies $\omega_{k}$ and eigenvectors
$\mathbf{u}(\omega_{k})$ whose site components are expressed, in bracket
notation, in terms of the site versors $\left\langle n\right\vert $ as%
\begin{equation}
u_{n}(\omega_{k})=\left\langle n\left\vert \varphi_{k}\right.  \right\rangle ,
\end{equation}
and the dynamical matrix is:%
\begin{equation}
\mathbb{M=}\left(
\begin{array}
[c]{cccc}%
\omega_{0}^{2}+\alpha\omega_{x}^{2} & -\alpha\omega_{x}^{2} & 0 & \cdots\\
-\omega_{x}^{2} & 2\omega_{x}^{2} & -\omega_{x}^{2} & \\
0 & -\omega_{x}^{2} & 2\omega_{x}^{2} & \\
\vdots &  &  & \ddots
\end{array}
\right)  .
\end{equation}

We define the\ Green's function operator:%
\begin{equation}
\mathbb{D}(\omega)=\left(  \omega^{2}\mathbb{I-M}\right)  ^{-1},
\label{eq_greenop}%
\end{equation}
whose matrix elements diverge when $\omega$ coincides with an eigenfrequency
$\omega_{k}$.

Consider an initial condition with all masses at equilibrium and an impulsive
force $m_{j}\dot{u}_{j}(0)\delta(t)$ applied\ to the $j$th mass, impossing an
initial velocity to it. In this case, the Green's function provides the time
evolution of the $i$th mass displacement:%
\begin{equation}
u_{i}(t)=D_{ij}(t)\dot{u}_{j}(0)=\frac{1}{2\pi}\int_{-\infty}^{\infty
}\mathrm{d}\omega e^{-\mathrm{i}\omega t}D_{ij}(\omega)\dot{u}_{j}(0),
\label{eq_displacement}%
\end{equation}
representing a position-velocity Green's function $D_{ij}(\omega)\equiv
D(u_{i},\dot{u}_{j};\omega)$, which is related to the\ position-position
Green's function:%
\begin{equation}
D(u_{i},u_{j};\omega)=-\mathrm{i}\omega D(u_{i},\dot{u}_{j};\omega).
\label{eq_standard Greens}%
\end{equation}

This gives the displacement of the $i$th mass when an initial displacement
$u_{j}(0)$ is imposed by the force:%
\begin{equation}
F_{j}(t)=m_{j}\frac{u_{j}(0)}{\tau}[\delta(t+\tfrac{1}{2}\tau)-\delta
(t-\tfrac{1}{2}\tau)]
\end{equation}
at the $j$th mass, where $\tau\ll1/\omega_{x}$.

In this work, we focus on the surface oscillator through $D_{00}(t)$. If
$\mathbb{U}$ is the matrix that performs the change of basis from sites to
diagonal form in $\mathbb{M}$, one has:%
\begin{equation}
U_{ik}=\left\langle i\right\vert \left.  \varphi_{k}\right\rangle ,
\end{equation}
that is, the $k$th eigenmode projection over $i$th site. In this case one can
write,%
\begin{equation}
D_{00}(\omega)=\sum_{k}\frac{\left\vert \left\langle 0\right\vert \left.
\varphi_{k}\right\rangle \right\vert ^{2}}{\omega^{2}-\omega_{k}^{2}}.
\label{eq_basis}%
\end{equation}

By performing the analytical continuation $\omega\rightarrow\omega
+\mathrm{i}\delta$ for $D_{00}(\omega)$ and taking its imaginary component,
one obtains the spectral density associated to the surface site. In the case
where $\alpha\rightarrow0$ bulk modes and surface modes have no projection and
oscillations survive undamped.

$D_{00}(\omega)$ can be obtained through an infinite order perturbation theory
\cite{PM01} that accounts for the surface mode corrections due to the presence
of the neighbor oscillators. Let us begin with the surface oscillator and a
single bulk mass. If the two masses are uncoupled ($\omega_{x}=0$) the surface
frequency is simply $\omega_{0}$ but if $\omega_{x}\neq0$,
Eq.(\ref{eq_greenop}) yields%
\begin{equation}
D_{00}(\omega)=\frac{1}{\omega^{2}-\omega_{0}^{2}-\alpha\omega_{x}^{2}%
-\alpha\dfrac{\omega_{x}^{4}}{\omega^{2}-\omega_{x}^{2}}}. \label{eq_two}%
\end{equation}

Here, $\omega_{0}$ is affected by the static correction $\alpha\omega_{x}^{2}$
and the dynamic correction $\alpha\Pi(\omega)$ due to presence of the other oscillator.

By taking the limit of the oscillators number to infinite, the dynamic
correction becomes%
\begin{equation}
\Pi(\omega)=\frac{\omega_{x}^{4}}{\omega^{2}-2\omega_{x}^{2}-\Pi(\omega)}.
\label{eq_correction}%
\end{equation}

Because of the presence of an infinite number of oscillators at the right, a
correction on the $n$th oscillator is just the same as that in the $(n+1)$th
oscillator. The solution of Eq.(\ref{eq_correction}) is complex and its
imaginary part gives the decay rate of a surface excitation with frequency
$\omega$. Therefore, in the thermodynamic limit, the temporal recurrences
\cite{Blaise} (Mesoscopic Echoes \cite{Mesoscopic}) disappear.

\section{Surface mode frequency}

In the case of a surface oscillator with natural frequency $\omega_{0}$,
coupled to an infinite number of bulk oscillators, the Green's function is:%
\begin{equation}
D_{00}(\omega)=\frac{1}{\omega^{2}-\omega_{0}^{2}-\alpha\omega_{x}^{2}%
-\alpha\Pi(\omega)}.
\end{equation}

Hence, in the region of continuous modes ($\left\vert \omega\right\vert
\leq2\omega_{x}$) one has%
\begin{equation}
D_{00}(\omega)=\frac{1}{1-\frac{\alpha}{2}}\frac{1}{\omega^{2}-\bar{\omega
}_{0}^{2}+\mathrm{i}\eta(\omega)\omega}, \label{eq_green}%
\end{equation}
where $\bar{\omega}_{0}=\omega_{0}/\sqrt{1-\alpha/2}$ is a first approximation
to the resonance frequency and%
\begin{equation}
\eta(\omega)=\frac{\alpha\omega_{x}}{1-\frac{\alpha}{2}}\sqrt{1-\left(
\frac{\omega}{2\omega_{x}}\right)  ^{2}}%
\end{equation}
describes the dissipation process. Comparing Eq.(\ref{eq_green}) with a
standard damped oscillator (frictional force proportional to velocity), it is
easy to see that in the broadband limit ($\alpha\rightarrow0,$ $\omega
_{x}\rightarrow\infty$ and $\alpha\omega_{x}=\mathrm{const}$.) they have the
same behavior taking $\eta_{0}\equiv\alpha\omega_{x}$ as friction coefficient.

For a dynamic description of the surface oscillator, we look for the pole
structure of $D_{00}(\omega)$, equivalent to find $\omega$ such that:%
\begin{equation}
\omega^{2}-\bar{\omega}_{0}^{2}+\mathrm{i}\eta(\omega)\omega=0.
\label{eq_polos1}%
\end{equation}

\begin{figure}
[h]
\begin{center}
\includegraphics*[
height=7.6926cm,
width=8.8041cm
]%
{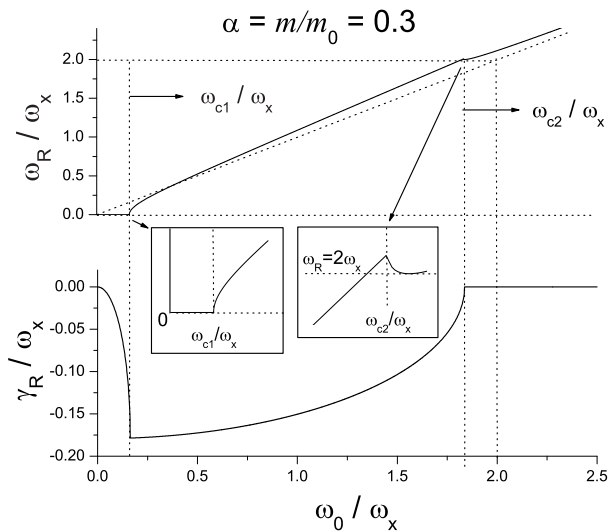}%
\caption{Surface Green's function pole. Above: real part (resonance). Below:
imaginary part (damping). The case $\omega_{R}=\omega_{0}$ is shown in dotted
line. Insets: overdamped - oscillatory transition (left), oscillatory -
localized (right).}%
\label{fig_resonance}%
\end{center}
\end{figure}

If we denote the surface mode frequency as $\tilde{\omega}_{R}=\omega
_{R}+\mathrm{i}\gamma_{R}$ and $p$ as%
\begin{equation}
p=\frac{\omega_{0}-\omega_{x}}{\omega_{x}\sqrt{1-\alpha}},
\end{equation}
then one has%
\[
\tilde{\omega}_{R}=\left\{
\begin{array}
[c]{cc}%
-\sqrt{\omega_{0}^{2}+\delta\left[  \omega_{-}^{2}+\sqrt{\omega_{+}%
^{4}-4\omega_{x}^{2}\omega_{0}^{2}}\right]  } & p<-1\\
\sqrt{\omega_{0}^{2}+\delta\left[  \omega_{-}^{2}-\mathrm{i}\sqrt{4\omega
_{x}^{2}\omega_{0}^{2}-\omega_{+}^{4}}\right]  } & \left\vert p\right\vert
\leq1\\
\sqrt{\omega_{0}^{2}+\delta\left[  \omega_{-}^{2}-\sqrt{\omega_{+}^{4}%
-4\omega_{x}^{2}\omega_{0}^{2}}\right]  } & p>1
\end{array}
\right.
\]
where $\omega_{\mathrm{\pm}}^{2}=\omega_{0}^{2}\pm\alpha\omega_{x}^{2}$ and
$\delta=\alpha/2(1-\alpha)$. The dependence of the pole with $\omega_{0}$ can
be seen in Fig.\ref{fig_resonance}.%

Here, we show the real and imaginary parts of the pole separately and we find
three well-defined regimes that emphasize the onset of quite different
dynamical properties.

\section{Dynamical phases}

\textbf{Localized - extended transition. }This transition occurs when
$\omega_{0}$ lays close to the upper band edge. There, the real part of the
pole makes an excursion into the band gap conserving an imaginary component
which is lost at the cusp zoomed in the inset. This is analogous to the
virtual states described by Hogreve \cite{Hogreve} for electronic states which
are missed under other approximations \cite{Economou, Desjonqueres}. They
become real normalized states which are exponentially localized \cite{Taylor}
when the pole touches back the band edge.

It can be seen in Fig.\ref{fig_resonance} that for large values of surface
natural frequency $(\omega_{0}\gg2\omega_{x})$ we only have real component in
$\tilde{\omega}_{R}$ and the oscillation frequency is approximately
$\omega_{0}$. In this localized regime, displacements are exponentially
smaller as the oscillators increase their distance to the surface. The lack of
an imaginary part implies that the displacement amplitude, independently of
bulk oscillators,\ survives indefinitely. In other words, there is no energy
propagation to bulk oscillators.

As $\omega_{0}$ decreases, below a critical frequency $\omega_{c2}=\omega
_{x}(1+\sqrt{1-\alpha})$, $\tilde{\omega}_{R}$ becomes complex. Its real part
is the resonance frequency whereas its imaginary part describes the
dissipation. This is the extended oscillatory regime where surface
oscillations have an $\omega_{R}$ frequency and the amplitude of displacement
decays with a lifetime proportional to $1/\gamma_{R}$.

\textbf{Oscillatory - overdamped transition.} For $\omega_{0}\ll\omega_{x}$
the system is in the regime where $\eta(\omega)\simeq\alpha\omega
_{x}/(1-\alpha/2)$ and hence Eq.(\ref{eq_polos1}) yields Ohmic dissipation in
which the surface kinetic energy decays into bulk modes without return. This
describes a frictional force proportional to the velocity. As $\omega_{0}$
goes below the critical frequency $\omega_{c1}=\omega_{x}(1-\sqrt{1-\alpha})$,
the pole of the Green's function $\tilde{\omega}_{R}$ becomes purely
imaginary. This is the overdamped regime where no oscillations occur. Notice
that $\gamma_{R}$ decreases at both sides of the transition, this means that a
further decrease of $\omega_{0}$ or increase in the friction $\eta$ also
implies a \textit{decrease} in the relaxation rate. All this counter intuitive
effect arrises from the exact solution of the simple mechanical model.

\begin{figure}
[h]
\begin{center}
\includegraphics*[
height=6.1264cm,
width=8.8041cm
]%
{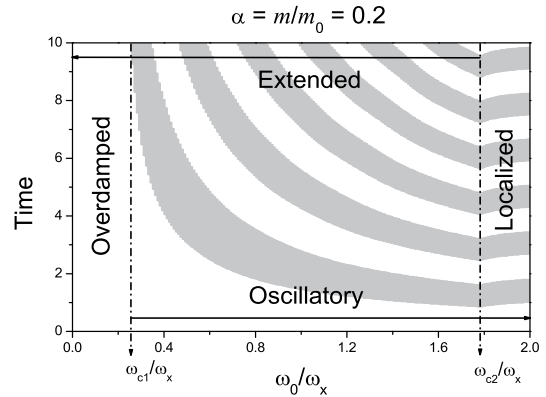}%
\caption{Evolution of the surface displacement. White spaces denote
$u_{0}(t)>0$ whereas gray spaces denote $u_{0}(t)<0$.}%
\label{fig_dynamic}%
\end{center}
\end{figure}

Considering Eq.(\ref{eq_displacement}), we can complete our previous analysis
obtaining the time evolution of surface displacement by performing a numerical
Fourier transform of the analytical expression for $D_{00}(\omega)$.\ These
results coincide with the numerical integration of the equation of motion in
the Hamilton-Jacobi representation which is performed with an \textit{ad hoc}
version of the Trotter-Suzuki method \cite{Trotter-Suzuki}. Setting the ratio
$\alpha$ between masses and Fourier transforming for several $\omega_{0}$
values in the previously mentioned regimes, we obtain the displacement pattern
shown in Fig.\ref{fig_dynamic}.%

Here, the time evolution of the surface displacement depends on the initial
condition. It is easy to see in this picture how gray regions "diverge" near
the critical frequency $\omega_{c1}$ so that for smaller $\omega_{0}$ values
$u_{0}(t)$ is always positive. On the other hand, the period of the
oscillation (reflected from the fringe edges) shows a cusp at $\omega_{c2}$
where it start to increase again. This is consistent with the decrease of
$\omega_{R}$ as $\omega_{0}$ crosses $\omega_{c2}$ (see inset in
Fig.\ref{fig_resonance}).

\section{Conclusions}

Dynamical features in dissipation processes have been described by a simple
classical model that contains the essential properties of a surface oscillator
interacting with bulk vibrational modes. We arrive to results equivalent to
the phenomenological description of dissipation where these effects are
summarized in a single term proportional to the oscillator velocity. A
complete analytical solution of the dynamics allowed us to identify a variety
of dynamical phases available for the vibrational excitations: localized,
extended oscillatory and overdamped. Two numerical methods (FFT and a
Trotter-Suzuki algorithm) have been employed to obtain the dynamics with
equivalent results.

The identification of the different dynamical regimes becomes very important
in nano-physics when cantilevers \cite{Cantilever} involving a relative small
number of atoms fail to satisfy the standard thermodynamical approximations.
Localization, weak diffusion and recurrences become usual phenomena in the
nanomechanical devices. In a quantum description of finite size systems, we
should consider the quantization of oscillation amplitudes. Our formalism,
based on the Green's functions in discrete but open systems, seems to be the
ideal tool for this purpose.

\end{document}